\documentstyle[12pt]{article}

\begin{document}

\title{A General Analysis of Corrections \\to the Standard See-saw 
Formula \\in Grand Unified Models}

\author{S.M. Barr and Bumseok Kyae \\Bartol Research Institute\\
University of Delaware\\Newark, DE 19716}

\date{}
\maketitle

\begin{abstract}

In realistic grand unified models there are typically extra 
vectorlike matter multiplets at the GUT scale that are needed to explain the
family hierarchy. These contain neutrinos that, when integrated out,
can modify the usual neutrino see-saw formula. A general analysis is
given. It is noted that such modifications can explain why the neutrinos
do not exhibit a strong family hierarchy like the other types of fermions.

\end{abstract}

\section{Introduction}

It is well-known that in simple grand unified theories (GUTs)
the masses of the light neutrinos are given by the see-saw formula
\begin{equation}
M_{\nu \nu} = - m_{\nu} {\cal M}^{-1} m_{\nu}^T,
\end{equation}
\noindent
where $m_{\nu}$ is the Dirac mass matrix connecting left- and right-handed
neutrinos and ${\cal M}$ is the Majorana mass matrix of the right-handed
neutrinos \cite{seesaw}. 
Typically, the Dirac matrix $m_{\nu}$ is closely related by the
unified symmetries to the Dirac mass matrices of the quarks and charged 
leptons, and like them exhibits a ``hierarchical" pattern.  This presents
something of a puzzle, since the evidence suggests that $M_{\nu \nu}$ does
not have a strongly hierarchical structure. (In particular, the neutrino
mixing angles $\theta_{atm}$ and $\theta_{sol}$ are not small, and the mass
ratio $\sqrt{\delta m^2_{sol}}/\sqrt{\delta m^2_{atm}}$ is about 0.2 \cite{atm}
\cite{sol}.)
One way to reconcile these facts is to posit a hierarchy in ${\cal M}$ that is 
just such as to nearly cancel the hierarchy in $m_{\nu}$. However, this idea 
seems contrived and is not easy to implement in a convincing way.

Another possibility is that $M_{\nu \nu}$ receives other contributions besides
the usual see-saw term given in Eq. (1). This can happen if the Higgs 
multiplets that break $B-L$ at the GUT scale and generate ${\cal M}$
also contain VEVs that break the weak interaction $SU(2)$. This leads to the 
so-called type II \cite{type2} and type III \cite{type3}
see-saw contributions. In this paper, we point
out a different kind of modification of the see-saw formula that comes from 
the existence of the superheavy vectorlike fermion multiplets that are quite a
typical feature of realistic grand unified models. We give a general 
analysis of the form of $M_{\nu \nu}$ in grand unified theories, showing under 
what conditions the different kinds of terms arise.

The standard analysis that leads to Eq. (1) considers the case where there is a left-handed neutrino ($\nu_i$) and a right-handed neutrino ($N^c_i$) for each
family ($i = 1,2,3)$. We will be interested in unified groups for which $\nu_i$
and $N^c_i$ are contained in a single irreducible multiplet ${\bf F}_i$. 
An obvious
example is $SO(10)$, where $\nu_i$ and $N^c_i$ are contained in the spinor
${\bf 16}_i$. In fact, we shall carry out the analysis in $SO(10)$, though it
will apply more generally. The crucial point that makes it necessary to 
go beyond the standard analysis is that in realistic unified models
there are usually {\it additional} multiplets of quarks and leptons besides the three familes ${\bf F}_i$. A very strong reason to suppose that extra 
fermions exist is the need to explain the hierarchical pattern of the mass 
matrices of the quarks and leptons, for the small entries of these matrices are generally assumed to arise from higher-dimensional effective operators 
obtained by integrating out such extra fermions at the GUT scale \cite{extraf}. In section 2 we will review the aspects of this idea that are relevant to 
our subsequent analysis. 

Since such extra fermions must get superlarge masses,
they necessarily are contained in a real (or ``vectorlike") set of multiplets, 
i.e. in real irreducible multiplets and/or in conjugate pairs of complex
multiplets of the unified group. For example, in $SO(10)$ they can be contained
in vector and tensor multiplets (${\bf 10}$, ${\bf 45}$, etc.) or in 
conjugate pairs of spinor and antispinor (${\bf 16} + \overline{{\bf 16}}$, 
etc.). Let us call this real set of multiplets ${\bf R}$. Thus the total quark 
and lepton content is contained in ${\bf F}_{i=1,2,3} + {\bf R}$. Typically,
one expects that there will be in ${\bf R}$ both left-handed neutrinos plus
their conjugates (which we shall denote by $\nu'_a + \overline{\nu}'_a$)
and right-handed neutrinos plus their conjugates (which we shall denote
by $N^{c \prime}_a + \overline{N^c}'_a$). Clearly the general
situation for neutrino masses becomes much more complicated than usually 
assumed. Nevertheless, a general analysis is possible that leads to some
simple conclusions. It can be shown under what conditions 
corrections to the usual 
see-saw formula arise, and what their form is. This analysis will be 
given in section 3. Our key results are contained in Eqs. (14), (19), and (23).

In section 4 we will present some simple illustrations of our general 
analysis in models taken from the literature. We will also show in
these examples how the extra terms in $M_{\nu \nu}$ beyond the usual
see-saw formula of Eq. (1) can resolve the puzzle of why $M_{\nu \nu}$
is not hierarchical even though $m_{\nu}$ probably is. In section 5, 
we give our conclusions and also discuss the similarities
and differences with the recent work of Nir and
Shadmi in \cite{nirshadmi}.

\section{Effective Dirac Mass Matrices and Hierarchies}

Before we turn to the problem of neutrino mass let us review how the effective
mass matrices of the three observed families of 
quarks and charged leptons arise from integrating
out superheavy vectorlike fermions. What we shall call the ``effective Dirac 
mass matrix of the neutrinos", which is important for our later
analysis, arises in exactly the same way. This effective Dirac mass matrix 
of the neutrinos is closely related by group theory in most unified theories 
to the effective mass matrices of the quarks and charged leptons, and thus 
shares their hierarchical structure. 

Consider fermions of type $f$ and $f^c$, where $f$ can be $u$, $d$, $\ell^-$,
or $\nu$, and $f^c$ can be $u^c$, $d^c$, $\ell^+$, or $N^c$. The $f$ are
weak-$SU(2)$ doublets and the $f^c$ are singlets. Imagine that in addition to 
the three families 
$f_i$ and $f^c_i$ ($i=1,2,3$) there are $f'_a + \overline{f}'_a$ ($a=1,...,
n_f$) and $f^{c \prime}_a + \overline{f^c}'_a$ ($a=1,...,n_{f^c}$) 
coming from ${\bf R}$.
The $f'_a$ and $f^{c \prime}_a$ have the same standard model quantum numbers as
the $f_i$ and $f^c_i$, while the $\overline{f}'_a$ and $\overline{f^c}'_a$ have opposite standard model quantum numbers. Then we can have superheavy mass terms of the form 
\begin{equation}
(f^T, f^{\prime T}) \left( \begin{array}{c} \hat{M}_f \\ M_f \end{array}
\right) \overline{f}' + (f^{c T}, f^{c \prime T}) \left( \begin{array}{c}
\hat{M}_{f^c} \\ M_{f^c} \end{array} \right) \overline{f^c}',
\end{equation}
\noindent
where we have suppressed indices. The matrix $M_f$ is $n_f \times n_f$; 
$\hat{M}_f$ is $3 \times n_f$; $M_{f^c}$ is $n_{f^c} \times n_{f^c}$; and
$\hat{M}_{f^c}$ is $3 \times n_{f^c}$. 
It is important to note that these matrices can depend on the superheavy VEVs 
of adjoint and other non-singlet Higgs fields.
There can also be, in the most general case, 
weak-$SU(2)$-breaking masses of the form
\begin{equation}
f^T m_f f^c + f^T m'_f f^{c \prime} +
f^{\prime T} m^{\prime \prime}_f f^{c \prime} + 
f^{\prime T} m^{\prime \prime \prime}_f f^c.
\end{equation}
\noindent
To find the effective low-energy
mass matrix, we must identify the light states, which we shall call
$f_{\ell}$ and $f^c_{\ell}$, and the superheavy states, which we shall call
$f_h$ and $f^c_h$. This can be done by unitary transformations
$(f^T, f^{\prime T}) = (f_{\ell}^T, f_h^T) {\cal U}_f$ and
$(f^{c T}, f^{c \prime T}) = (f_{\ell}^{c T}, f_h^{c T}) {\cal U}_{f^c}$, 
where
\begin{equation}
{\cal U}_K \left( \begin{array}{c} \hat{M}_K \\ M_K \end{array} \right)
= \left( \begin{array}{c} 0 \\ \overline{M}_K \end{array} \right), \; 
K = f,f^c.
\end{equation}
\noindent
Then
\begin{equation}
{\cal U}_K = \left( \begin{array}{cc} U_K & -U_K Z_K \\
Z_K^{\dag} U_K & V_K \end{array} \right),
\end{equation}
\noindent
where $Z_K = \hat{M}_K M^{-1}_K$, $U_K = (I + Z_K Z_K^{\dag})^{-1/2}$,
$V_K = (I + Z_K^{\dag} Z_K)^{-1/2}$,
and $\overline{M}_K = V_K^{-1} M_K$. This transformation allows us to write
$f = U_f^T f_{\ell} + ...$, $f' = - Z_f^T U^T_f f_{\ell} 
+ ...$, $f^c = U_{f^c}^T f^c_{\ell} + ...$, and $f^{c \prime} =
- Z_{f^c}^T U^T_{f^c} f^c_{\ell} + ...$, where the dots
represent terms with heavy eigenstates. Then, inserting these results into
Eq. (3), we may write the effective mass 
term of the light eigenstates as $f^T_{\ell} \overline{m}_f f^c_{\ell}$, 
where 
\begin{equation}
\overline{m}_f = U_f \left( m_f - m'_f Z_{f^c}^T
- Z_f m^{\prime \prime \prime}_f +
Z_f m^{\prime \prime}_f Z_{f^c}^T
\right) U^T_{f^c}.
\end{equation}
\noindent
The four terms in this expression have the diagrammatic interpretations 
shown in Fig. 1.

If the mixings are small, then $|Z_K| \ll 1$, and $U_K \cong I$. The first 
term in the effective 
mass matrix $\overline{m}_f$ is then $\cong m_f$, which just
comes from the renormalizable operator corresponding to 
Fig. 1(a). This contribution is usually assumed to give the largest elements
in the quark and lepton mass matrices (e.g. the 33 elements). The other
terms in $\overline{m}_f$ come from the higher-order effective operators
corresponding to the other diagrams in Fig. 1. These are usually assumed to
give the smaller elements in the mass matrices. If, for instance, the
elements of $\hat{M}_K$ were of order $\epsilon$ times the elements of $M_K$,
then the higher order terms in $\overline{m}_f$ would be suppressed by powers 
of $\epsilon$ relative to the lowest order terms. All models in which the
``textures" of mass matrices are explained by tree-level diagrams fit the 
general scheme we have just outlined. We will consider some realistic
examples taken from the literature in section 4. 

It is usually assumed that the term $m_f$ has the form $diag(0,0,1)$ and
gives the masses of the third family. The matrices $M_K$ and/or 
$\hat{M}_K$ can contain VEVs of adjoints or larger representations of Higgs
fields, and thus introduce group generators into the smaller elements of
$\overline{m}_f$ which can explain the Georgi-Jarlskog factors and other
differences between the matrices of up quarks, down quarks, charged
leptons, and neutrinos. It is important to emphasize that the discussion
we have just given applies to the neutrinos as well as the other fermions, 
and that the
matrix $\overline{m}_{\nu}$ given by Eq. (6) with $f= \nu$ and $f^c = N^c$
is just the effective Dirac mass matrix of the
neutrinos. This is the matrix that naively would be expected to appear in the
see-saw formula of Eq. (1). Whether it does appear in that formula, and
whether there are corrections to it, can only be determined by analyzing
the entire neutrino mass matrix, including the $(B-L)$-violating terms that
give the Majorana mass matrix of the right-handed neutrinos $N^c$ and 
$N^{c \prime}$. This will be done in the next section.

\section{A General Analysis of Neutrino Masses} 

We will give the analysis in the language of $SO(10)$, though 
it is simple to generalize to larger groups simply by decomposing under the 
$SO(10)$ subgroup. We may
take ${\bf F}_i = {\bf 16}_i$. The real representations ${\bf R}$ of
quarks and leptons can consist of some number of ${\bf 1}, {\bf 10},
({\bf 16} + \overline{{\bf 16}}), {\bf 45}, {\bf 54}$, or even larger
multiplets (though it is hard to find models in the literature where larger
multiplets of quarks and leptons appear). Of the representations just
listed, $({\bf 16} + \overline{{\bf 16}})$ contains both singlet and doublet 
neutrinos and their conjugates; ${\bf 10}$
contains only doublet neutrinos and their conjugates; and  
${\bf 1}$, ${\bf 45}$, and ${\bf 54}$ contain only singlet neutrinos.
 
We will proceed in stages from the simplest situation to
the most complicated. In subsection 3.1, we assume that ${\bf R}$ contains
only spinor-antispinor pairs, which leads to a relatively simple set of 
possibilities. (It should be noted that
for such models the right-handed Majorana masses of the neutrinos must
come from terms of the form ${\bf 16} \; {\bf 16} \langle 
\overline{{\bf 126}}_H
\rangle$.) The models of Ref. \cite{adhrs}, for example,
fall within this class. In subsection 3.2, we 
allow ${\bf R}$ to contain both spinor-antispinor pairs and vectors. The
possibilities become slightly more complicated. The right-handed Majorana
neutrino masses must still come from ${\bf 16} \; {\bf 16} \langle
\overline{{\bf 126}}_H \rangle$. The model 
of Ref. \cite{bpw}, for example, falls
within this class. Finally, in subsection 3.3, we allow ${\bf R}$ to contain 
not only spinor-antispinor pairs and vectors but also the representations
${\bf 1}$, ${\bf 45}$ and ${\bf 54}$. The right-handed Majorana masses
of the neutrinos can then come either from either ${\bf 16} \; {\bf 16}
\langle \overline{{\bf 126}}_H \rangle$ or from non-renormalizable 
operators of the form
${\bf 16} \; {\bf 16} \; \langle \overline{{\bf 16}}_H \rangle
\langle \overline{{\bf 16}}_H \rangle/M_{GUT}$ obtained by integrating out
the singlet or tensor multiplets of fermions. The model of Ref. \cite{abb},
for example, falls in this wider class.

\subsection{$SO(10)$ models where ${\bf R}$ contains only spinor-antispinor 
pairs.}

We first consider models where ${\bf R}$ consists of $({\bf 16}_a +
\overline{{\bf 16}}_a)$, $a = 1, ..., N$.
In analyzing this case we will make one very weak assumption, namely that
there are no ${\bf 210}$ Higgs multiplets coupling
spinors to antispinors of quarks and leptons. This means that the only
Higgs multiplets that do couple spinors to antispinors will be in
${\bf 1}$ or ${\bf 45}$, neither of which contain $SU(2)$-doublet
components that can get VEVs.

In the case being considered, the neutrinos fall into just three groups:
$\nu_i, N^c_i \in {\bf 16}_i = {\bf F}_i$ ($i = 1,2,3$); $\nu'_a, 
N^{c \prime}_a \in {\bf 16}_a \in {\bf R}$ ($a = 1, ..,N$); and
$\overline{\nu}'_a, \overline{N^c}'_a \in \overline{{\bf 16}}_a \in 
{\bf R}$ ($a = 1, ...,N$). Given our weak assumption, the most general 
form of the mass matrix is (suppressing indices):
\begin{equation}
\left( \nu,\nu',\overline{\nu}',N^c,N^{c \prime}, \overline{N^c}' \right)
\left( \begin{array}{cccccc} \mu & \mu' & \hat{M}_{\nu} & m_{\nu} & m'_{\nu} 
& 0 \\
\mu^{\prime T} & \mu^{\prime \prime} & M_{\nu} & 
m^{\prime \prime \prime}_{\nu} & m^{\prime \prime}_{\nu}
& 0 \\ 
\hat{M}^T_{\nu} & M^T_{\nu} & - & 0 & 0 & - \\
m^T_{\nu} & m^{\prime \prime \prime T}_{\nu} & 0 & {\cal M} & {\cal M}' &
\hat{M}_{N^c} \\
m^{\prime T}_{\nu} & m^{\prime \prime T}_{\nu} & 0 & {\cal M}^{\prime T} &
{\cal M}^{\prime \prime} & M_{N^c} \\
0 & 0 & - & \hat{M}^T_{N^c} & M^T_{N^c} & \tilde{{\cal M}} 
\end{array} \right) \left( \begin{array}{c} \nu \\ \nu' \\ \overline{\nu}' \\
N^c \\ N^{c \prime} \\ \overline{N^c}' \end{array} \right).
\end{equation}
\noindent
The dashes represent entries that can be neglected because they lead to
contributions to $M_{\nu \nu}$ higher order than $M_W^2/M_{GUT}$. 
The matrices denoted with lower-case $m$ come from $SU(2)$-breaking vacuum
expectation values (VEVs) that couple ${\bf 16}$ to ${\bf 16}$ 
($m_{\nu}$, $m'_{\nu}$, $m^{\prime \prime}_{\nu}$, 
$m^{\prime \prime \prime}_{\nu}$). The matrices denoted by capital $M$
are GUT-scale masses that couple ${\bf 16}$ to
$\overline{{\bf 16}}$ ($M_{\nu}$, $\hat{M}_{\nu}$, $M_{N^c}$, $\hat{M}_{N^c}$).
The preceding matrices are of types considered in section 2. However,
the matrices denoted by calligraphic ${\cal M}$ were not considered in section 
2. They are the GUT-scale, $(B-L)$-violating ``right-handed Majorana mass 
matrices" and come from terms of the form  
${\bf 16} \; {\bf 16} \; \langle \overline{{\bf 126}}_H \rangle$
(for ${\cal M}$, ${\cal M}'$, ${\cal M}^{\prime \prime}$, 
${\cal M}^{\prime \prime \prime}$) or of the form $\overline{{\bf 16}} \;
\overline{{\bf 16}} \; \langle {\bf 126}_H \rangle$ (for $\tilde{{\cal M}}$).
The matrices denoted by $\mu$ are $O(M_W^2/M_{GUT})$ masses coming from
$SU(2)$-triplet VEVs. These (if they exist)
are the so-called type II see-saw contributions

If we set the $(B-L)$-violating masses (${\cal M}$ etc.) to zero, we will
end up with the situation already analyzed in section 2. That is, we
will end up (after integrating out superheavy fields) with three light 
$\nu$ and
three light $N^c$ connected by a weak-scale effective Dirac mass matrix
$\overline{m}_{\nu}$ given by Eq. (6) with $f= \nu$ and $f^c = N^c$. To see 
this, one simply transforms the full matrix in Eq. (7) (with ${\cal M}$ etc. 
set equal to zero) by
multiplying it from the left by ${\cal U}$
and from the right by ${\cal U}^T$, where
\begin{equation}
{\cal U} = \left( \begin{array}{cccccc} 
U_{\nu} & - U_{\nu} Z_{\nu} & 0 & 0 & 0 & 0 \\
Z_{\nu}^{\dag} U_{\nu} & V_{\nu} & 0 & 0 & 0 & 0 \\
0 & 0 & I & 0 & 0 & 0 \\ 
0 & 0 & 0 & U_{N^c} & - U_{N^c} Z_{N^c} & 0 \\ 
0 & 0 & 0 & Z_{N^c}^{\dag} U_{N^c} & V_{N^c} & 0 \\
0 & 0 & 0 & 0 & 0 & I \end{array} \right).
\end{equation}
\noindent
See Eqs. (4) and (5).
If one makes the same transformation {\it without} setting the
$(B-L)$-violating masses (${\cal M}$ etc.) to zero, the matrix in Eq. (7)  
takes the form
\begin{equation}
\left( \begin{array}{cccccc} \overline{\mu} & - & 0 & \overline{m}_{\nu} & 
\overline{m'}_{\nu} & 0 \\
- & - & \overline{M}_{\nu} & - & -
& 0 \\ 0 & \overline{M}^T_{\nu} & - & 0 & 0 & - \\
\overline{m}^T_{\nu} & - & 0 & \overline{{\cal M}} & \overline{{\cal M}'} &
0 \\
\overline{m'}^T_{\nu} & - & 0 & \overline{{\cal M}'}^T &
\overline{{\cal M}^{\prime \prime}} & \overline{M}_{N^c} \\
0 & 0 & - & 0 & \overline{M}^T_{N^c} & \tilde{{\cal M}} 
\end{array} \right).
\end{equation}
\noindent
Note that the matrices shown in the lower-right 
$5 \times 5$ block have GUT-scale
entries, while the matrices in the first row and column have entries
that are weak-scale or smaller.
The dashes represent matrices that can be neglected. 
The matrix $\overline{m}_{\nu}$
is the same as that given in Eq. (6) with $f = \nu$ and $f^c = N^c$: 
\begin{equation}
\overline{m}_{\nu} = U_{\nu} \left( m_{\nu} - m'_{\nu} Z_{N^c}^T
- Z_{\nu} m^{\prime \prime \prime}_{\nu} +
Z_{\nu} m^{\prime \prime}_{\nu} Z_{N^c}^T
\right) U^T_{N^c}.
\end{equation}
\noindent
That is,
it is just the ``effective Dirac neutrino mass matrix." The matrix 
$\overline{m'}_{\nu}$ is given by
\begin{equation}
\overline{m'}_{\nu} = U_{\nu} \left( m'_{\nu} + m_{\nu} Z_{N^c}^*
- Z_{\nu} m^{\prime \prime}_{\nu}
- Z_{\nu} m^{\prime \prime \prime} Z_{N^c}^* \right) V^T_{N^c}.
\end{equation}
\noindent
For later reference, we give here the expressions for the remaining barred
matrices: $\overline{{\cal M}} = U_{N^c}({\cal M} - {\cal M}' Z^T_{N^c}
- Z_{N^c} {\cal M}^{\prime T} + Z_{N^c} {\cal M}^{\prime \prime} Z^T_{N^c})
U^T_{N^c}$; $\overline{{\cal M}'} = U_{N^c}({\cal M}' + {\cal M} Z^*_{N^c}
- Z_{N^c} {\cal M}^{\prime \prime} - Z_{N^c} {\cal M}^{\prime T} Z^*_{N^c})
V^T_{N^c}$; $\overline{{\cal M}^{\prime \prime}} = 
V_{N^c}({\cal M}^{\prime \prime} + {\cal M}^{\prime T} Z^*_{N^c}
+ Z^{\dag}_{N^c} {\cal M}' + Z^{\dag}_{N^c} {\cal M} Z^*_{N^c})
V^T_{N^c}$.

From Eq. (9) we see that the effective mass matrix of the three light
neutrinos
is given by
\begin{equation}
M_{\nu \nu} =
- \left( \overline{m}_{\nu}, \overline{m'}_{\nu}, 0 \right)
\left( \begin{array}{ccc} \overline{{\cal M}} & \overline{{\cal M}'} & 0 \\
\overline{{\cal M}'}^T & \overline{{\cal M}^{\prime \prime}}^T &
\overline{M}_{N^c} \\ 0 & \overline{M}_{N^c}^T & \tilde{{\cal M}} \end{array}
\right)^{-1} \left( \begin{array}{c} \overline{m}_{\nu}^T \\
\overline{m'}_{\nu}^T \\ 0 \end{array} \right) \;\; +  \;\; \overline{\mu}.
\end{equation}
\noindent
Let us first consider an easy case.

{\it Case I:} $\tilde{{\cal M}} = 0$. (That is, there are no GUT-scale,
$(B-L)$-violating mass terms of the form 
$\overline{{\bf 16}} \; \overline{{\bf 16}}
\langle {\bf 126}_H \rangle$.)  In this case, the inverse of the $3 \times 3$ 
block matrix in Eq. (12) is easily seen to have vanishing ``12", ``21", and
``22" entries. So one obtains the standard see-saw formula together with
possible type II corrections:
\begin{equation}
M_{\nu \nu} = - \overline{m}_{\nu} \overline{{\cal M}}^{-1} 
\overline{m}_{\nu}^T \;\; + \;\; \overline{\mu}.
\end{equation}
\noindent
Note the important fact that not only does $M_{\nu \nu}$ have the standard
see-saw form, but the Dirac mass matrix $\overline{m}_{\nu}$ appearing
in the formula is just the effective Dirac mass matrix that one would
compute by setting the ``right-handed Majorana masses" (which 
violate $B-L$) to zero and integrating out the extra neutrinos and 
antineutrinos in ${\bf R}$. That is, it is the same Dirac neutrino
matrix that is related by the unified symmetries to the effective
mass matrices of the up quarks, down quarks, and charged leptons, and 
is like them hierarchical. In other words, in Case I the
naive procedure for computing $M_{\nu \nu}$ is correct.

{\it Case II:} $\tilde{{\cal M}} \neq 0.$ (That is, there do exist
$(B-L)$-violating terms of the form
$\overline{{\bf 16}} \; \overline{{\bf 16}}
\langle {\bf 126}_H \rangle$.) Here the inversion of the matrix in Eq. (12)
can still be done and yields the result
\begin{equation}
M_{\nu \nu}  =  - \overline{m}_{\nu} \overline{{\cal M}}^{-1} 
\overline{m}_{\nu}^T \; + \; \overline{\mu} \; - \;
m_* {\cal M}_*^{-1} m_*^T,
\end{equation}
\noindent
where
\begin{equation}
m_* \equiv \overline{m'}_{\nu} - \overline{m}_{\nu} \overline{{\cal M}}^{-1}
\overline{{\cal M}'},
\end{equation}
\noindent
and
\begin{equation}
{\cal M}_* \equiv \overline{{\cal M}^{\prime \prime}} - 
\overline{{\cal M}'}^T \overline{\cal M}^{-1} \overline{{\cal M}'} 
- \overline{M}_{N^c} \tilde{{\cal M}}^{-1} \overline{M}_{N^c}^T.
\end{equation}
\noindent
The three terms in Eq. (14) have the diagrammatic interpretation given in
Fig. 2. The third term of Eq. (14)   
looks complicated, but that is because we are considering very general 
possibilities. In realistic models, many of the submatrices in the mass matrix
in Eq. (7) vanish, so that Eqs. (15) and (16) 
often reduce to quite simple expressions,
as we shall see in section 4. We shall also see that the 
third term in Eq. (14) can explain why $M_{\nu \nu}$ is not hierarchical. 
Note that in
the limit $\tilde{{\cal M}} \longrightarrow 0$, one has ${\cal M}_*^{-1}
\longrightarrow 0$, so that one recovers Case I.

\subsection{$SO(10)$ models where ${\bf R}$ contains only spinor-antispinor
pairs and vectors.}

We now consider models where ${\bf R}$ consists of $({\bf 16}_a + 
\overline{{\bf 16}}_a)$, $a = 1, ..., N$, and ${\bf 10}_m$, $m = 1, ..., M$.
The vectors ${\bf 10}_m$ contain additional $\nu'$ and $\overline{\nu}'$.
There can be the following new kinds of terms: 
(a) $M_{mn} ({\bf 10}_m {\bf 10}_n)$. This gives new
contributions to the matrix we have 
been calling $M_{\nu}$. (b) $({\bf 10}_m {\bf 16}_{i,a}) \langle 
{\bf 16}_H \rangle$.
This can give new contributions to $M_{\nu}$ (or $\hat{M}_{\nu}$), if we take
the $\overline{\nu}'$ component of ${\bf 10}_m$ and the $\nu$ (or $\nu'$)
component of the ${\bf 16}_i$ (or ${\bf 16}_a$). It can also give weak-scale
masses connecting the $\overline{\nu}'$ component of ${\bf 10}_m$ to
the $N^c$ (or $N^{c \prime}$) component of the ${\bf 16}_i$ (or 
${\bf 16}_a$). However, since these lead to negligible contributions to 
$M_{\nu \nu}$, they are represented by dashes
below. (c) $({\bf 10}_m \overline{{\bf 16}}_a) \langle \overline{{\bf 16}}_H
\rangle$. This can give new contributions to $M_{\nu}$, but also a new
and significant kind of contribution, namely to the submatrix that couples
$\nu'$ in the vector(s) to the $\overline{N^c}'$ in the antispinor(s).
This entry will be denoted below by the matrix $\tilde{m}$. 
Altogether, then, the matrix has almost the same form as in Eq.(7):
\begin{equation} 
\left( \nu,\nu',\overline{\nu}',N^c,N^{c \prime}, \overline{N^c}' \right)
\left( \begin{array}{cccccc} \mu & \mu' & \hat{M}_{\nu} & m_{\nu} & m'_{\nu} 
& 0 \\
\mu^{\prime T} & \mu^{\prime \prime} & M_{\nu} & 
m^{\prime \prime \prime}_{\nu} & m^{\prime \prime}_{\nu}
& \tilde{m} \\ 
\hat{M}^T_{\nu} & M^T_{\nu} & - & - & - & - \\
m^T_{\nu} & m^{\prime \prime \prime T}_{\nu} & - & {\cal M} & {\cal M}' &
\hat{M}_{N^c} \\
m^{\prime T}_{\nu} & m^{\prime \prime T}_{\nu} & - & {\cal M}^{\prime T} &
{\cal M}^{\prime \prime} & M_{N^c} \\
0 & \tilde{m}^T & - & \hat{M}_{N^c} & M^T_{N^c} & \tilde{{\cal M}} 
\end{array} \right) \left( \begin{array}{c} \nu \\ \nu' \\ \overline{\nu}' \\
N^c \\ N^{c \prime} \\ \overline{N^c}' \end{array} \right).
\end{equation}
\noindent
If the matrix we are calling $\tilde{m}$ vanishes, then the form of this
matrix is not different, except for negligible entries, from the matrix
we considered in the previous subsection. So the results from that analysis
still apply, including Eqs. (13) and (14). However, if $\tilde{m} \neq 0$,
there are new cases to consider.

{\it Case III:} $\tilde{m} \neq 0$ and $\tilde{{\cal M}} \neq 0$. The 
general expression for $M_{\nu \nu}$ is easily found to be
\begin{equation}
M_{\nu \nu} =
- \left( \overline{m}_{\nu}, \overline{m'}_{\nu}, -U_{\nu} Z_{\nu} \tilde{m} 
\right)
\left( \begin{array}{ccc} \overline{{\cal M}} & \overline{{\cal M}'} & 0 \\
\overline{{\cal M}'}^T & \overline{{\cal M}^{\prime \prime}}^T &
\overline{M}_{N^c} \\ 0 & \overline{M}_{N^c}^T & \tilde{{\cal M}} \end{array}
\right)^{-1} \left( \begin{array}{c} \overline{m}_{\nu}^T \\
\overline{m'}_{\nu}^T \\ -\tilde{m}^T Z^T_{\nu} U^T_{\nu} 
\end{array} \right) \;\; +  \;\; \overline{\mu}.
\end{equation}
\noindent
This case is very complicated in general. However, there is a
subcase that is relatively simple, namely if the elements of 
$\overline{M}_{N^c}$ are very
small compared to the $(B-L)$-violating entries ($\tilde{{\cal M}}$, 
${\cal M}$, etc.) then we can approximate the ``33" component
of the inverse matrix in Eq. (18) by $\tilde{{\cal M}}^{-1}$, giving
\begin{equation}
\begin{array}{ccl}
M_{\nu \nu} & \cong & - \overline{m}_{\nu} \overline{{\cal M}}^{-1} 
\overline{m}_{\nu}^T + \overline{\mu} - m_* {\cal M}_*^{-1} m_*^T \\ \\
 & - & U_{\nu} \left( \hat{M}_{\nu} M_{\nu}^{-1} \tilde{m} 
\tilde{{\cal M}}^{-1} \tilde{m}^T M_{\nu}^{T -1} \hat{M}_{\nu}^T \right)
U^T_{\nu}, \end{array}
\end{equation}
\noindent
Where the first three terms are the same as in Eq. (14), and the quantities
with asterisk subscripts are still defined by Eqs. (15) and (16). The last term in Eq. (19) is the contribution from $\tilde{m}$. 

\subsection{$SO(10)$ models where ${\bf R}$ contains spinor-antispinor
pairs, vectors, singlets and tensors}

We now introduce, in addition to spinor-antispinor pairs $({\bf 16} +
\overline{{\bf 16}})$ and vectors ${\bf 10}$, $SO(10)$ singlet and tensor 
multiplets. We shall only allow the rank-2 tensors ${\bf 45}$ and ${\bf 54}$. 
(It is unusual to find multiplets of quarks and leptons larger than these in
published models.) The ${\bf 1}$, ${\bf 45}$, and ${\bf 54}$
contain leptons that are singlets
under the standard model group, which we shall denote by $S_I$. The
existence of at least three such singlets allows the 
``right-handed Majorana mass matrix" of the neutrinos
to be generated by effective operators of the form  ${\bf 16} \; {\bf 16}
\langle \overline{{\bf 16}}_H \rangle \langle \overline{{\bf 16}}_H
\rangle/M_{GUT}$, which come from integrating the singlets out. The 
existence of operators of the form ${\bf 16} \; {\bf 16} \langle 
\overline{{\bf 126}}_H \rangle$ would therefore be unnecessary.

The general case of neutrino mass where ${\bf R}$ contains 
spinor-antispinor pairs, vectors, singlets and small tensors of $SO(10)$ 
is very complicated. So we shall restrict our attention to a subcase defined
by the following assumptions: (1) There are exactly three
singlets $S_I$. (2) There are no $\overline{{\bf 126}}$ or
${\bf 126}$ of Higgs fields. (3) There are no $SU(2)$-breaking
masses from terms of the form ${\bf 10}_m \overline{{\bf 16}}_a
\langle \overline{{\bf 16}}_H \rangle$. i.e. in the notation of the
previous subsection, $\tilde{m} = 0$.

With these assumptions, the following new types of terms are allowed:
(a) $M {\bf 1} \; {\bf 1}$, $M {\bf 45} \; {\bf 45}$, and $M {\bf 54} 
\; {\bf 54}$. These lead to the mass matrix $M_S$ below.
(b) ${\bf 10} \; {\bf 1} \langle {\bf 10}_H \rangle$, and similar terms with 
the singlet replaced by a tensor. These couple $S$ to $\nu'$ and 
$\overline{\nu}'$, giving the entries denoted $g'$ and $\tilde{g}$ below.
(c) ${\bf 16} \; {\bf 1} \langle \overline{{\bf 16}}_H \rangle$, and similar
terms with the singlet replaced by a tensor. These couple
$S$ to $\nu$, $N^c$, $\nu'$, and $N^{c \prime}$, giving the entries denoted
$g$, $G$, $g'$, and $G'$ below. (d) $\overline{{\bf 16}} \; {\bf 1}
\langle {\bf 16}_H \rangle$, and similar terms with the singlet replaced by a 
tensor. These couple $S$ to $\overline{\nu}'$ and $\overline{N^c}'$, giving
the terms denoted by $\tilde{g}$ and $\tilde{G}$ below.

Given assumption (2), the kinds of masses we denoted above by 
${\cal M}$, ${\cal M}'$, ${\cal M}^{\prime \prime}$, and $\tilde{{\cal M}}$
all vanish, as do those we denoted by $\mu$, $\mu'$, etc. 
The general form of the mass matrix 
consistent with these assumptions is thus
\begin{equation}
\left( \nu,\nu',\overline{\nu}',N^c,N^{c \prime}, \overline{N^c}', S \right)
\left( \begin{array}{ccccccc} 0 & 0 & \hat{M}_{\nu} & m_{\nu} & m'_{\nu} 
& 0 & g \\
0 & 0 & M_{\nu} & 
m^{\prime \prime \prime}_{\nu} & m^{\prime \prime}_{\nu}
& 0 & g' \\ 
\hat{M}^T_{\nu} & M^T_{\nu} & 0 & 0 & 0 & - & \tilde{g} \\
m^T_{\nu} & m^{\prime \prime \prime T}_{\nu} & 0 & 0 & 0 &
\hat{M}_{N^c} & G \\
m^{\prime T}_{\nu} & m^{\prime \prime T}_{\nu} & 0 & 0 &
0 & M_{N^c} & G' \\
0 & 0 & - & \hat{M}^T_{N^c} & M^T_{N^c} & 0 & \tilde{G} \\
g^T & g^{\prime T} & \tilde{g}^T & G^T & G^{\prime T} & \tilde{G}^T & M_S 
\end{array} \right) \left( \begin{array}{c} \nu \\ \nu' \\ \overline{\nu}' \\
N^c \\ N^{c \prime} \\ \overline{N^c}' \\ S \end{array} \right).
\end{equation}
\noindent
Rotating the basis by the same unitary transformation in Eq. (8) (leaving
the singlets $S$ alone) one obtains
the matrix 
\begin{equation}
\left( \begin{array}{ccccccc} 0 & 0 & 0 & \overline{m}_{\nu} & 
\overline{m'}_{\nu} & 0 & \overline{g} \\
0 & 0 & \overline{M}_{\nu} & - & - & 0 & - \\ 
0 & \overline{M}^T_{\nu} & 0 & 0 & 0 & - & - \\
\overline{m}^T_{\nu} & - & 0 & 0 & 0 & 0 & \overline{G} \\
\overline{m'}^T_{\nu} & - & 0 & 0 & 0 & M_{N^c} & \overline{G'} \\
0 & 0 & - & 0 & \overline{M}^T_{N^c} & 0 & \tilde{G} \\
\overline{g}^T & - & - & \overline{G}^T & \overline{G'}^T & \tilde{G}^T & M_S 
\end{array} \right).
\end{equation}
\noindent
The light neutrino mass matrix is thus given by
\begin{equation}
M_{\nu \nu} = - (\overline{m}_{\nu}, \overline{m'}_{\nu}, 0, \overline{g})
\left( \begin{array}{cccc} 0 & 0 & 0 & \overline{G} \\
0 & 0 & \overline{M}_{N^c} & \overline{G'} \\
0 & \overline{M}_{N^c}^T & 0 & \tilde{G} \\
\overline{G}^T & \overline{G'}^T & \tilde{G}^T & M_S \end{array}
\right)^{-1} \left( \begin{array}{c} \overline{m}_{\nu}^T \\
\overline{m'}_{\nu}^T \\ 0 \\ \overline{g}^T \end{array} \right),
\end{equation}
\noindent
which yields
\begin{equation}
M_{\nu \nu} = - \overline{m}_{\nu} {\cal M}_{eff}^{-1} \overline{m}_{\nu}^T
- (\overline{m}_{\nu} \; A^T + A \; \overline{m}_{\nu}^T),
\end{equation}
\noindent
where
\begin{equation}
\begin{array}{ccl}
{\cal M}_{eff} & \equiv & - \overline{G}^{T -1}( M_S 
- \overline{G'}^T \overline{M}_{N^c}^{T -1} \tilde{G} 
- \tilde{G}^T \overline{M}_{N^c}^{-1} \overline{G'}) \overline{G}^{-1} \\ 
& & \\
A & \equiv & \overline{g} \overline{G}^{-1} 
- \overline{m'}_{\nu} \overline{M}_{N^c}^{T -1} \tilde{G} \overline{G}^{-1}.
\end{array}
\end{equation}
\noindent
The first term in Eq. (23) is just the standard (or ``type I")
see-saw contribution. The remaining terms in Eq. (23) have the so-called 
``type III" see-saw form. The first term in $A$ is, in fact, exactly the
type III contribution discussed in Ref. \cite{type3}, 
while the second term in $A$
is an additional contribution coming from the presence of the extra
spinor-antispinor pairs and vectors.

By relaxing some of the assumptions we made at the beginning of this 
subsection, one can have more complicated situations, in which there are
contributions of all types: type I, type II, type III, and the additional 
contributions coming from $\tilde{{\cal M}} \neq 0$ and $\tilde{m} \neq 0$
discussed previously. However, in most models one does not have the most
general possible situation.

\section{Examples from Realistic Published Models}

\subsection{A model where ${\bf R}$ contains only spinor-antispinor pairs}

In Ref. \cite{adhrs} an approach to understanding quark and lepton mass
hierarchies is proposed that is based on $SO(10)$ and the Froggatt-Nielson
idea \cite{fn}. A family $U(1)$ is introduced that is broken by the VEVs 
of familons.
These familons are in singlets or adjoints of the $SO(10)$. There are
extra vectorlike multiplets of quarks and leptons consisting of equal
numbers of spinors and antispinors. When these superheavy spinors and 
antispinors are integrated out, one is left with effective Yukawa operators 
that involve various powers of the familon field VEVs. The operators that 
are of higher order in these VEVs are more suppressed, giving rise to a 
hierarchy of masses. 

In Ref. \cite{adhrs} a set of criteria is imposed that leads to nine 
realistic models
of the type just described. These models are specified in that paper by the 
form of the effective Yukawa operators. The details of how these operators
arise are not described in detail in that paper, except in the case of
``model 9". (See Appendix A of Ref. \cite{adhrs}.) We shall take as our example a simplified version of model 9, in which the operators that contribute to 
the masses of the lightest family (and the fields that must be integrated 
out to produce these operators) are neglected. This simplified version 
will be sufficient for our purposes.

The quark and lepton content of the model is the following, with the family
charges given in parentheses: 
${\bf 16}_1 (2)$, ${\bf 16}_2 (1)$, ${\bf 16}_3 (0)$, 
${\bf 16}'_1 (-1)$, ${\bf 16}'_2 (-\frac{1}{2})$, 
$\overline{{\bf 16}}'_1 (\frac{3}{2})$,
$\overline{{\bf 16}}'_2 (1)$. The weak $SU(2)_L \times U(1)_Y$ 
breaking is done by a vector: ${\bf 10}_H$. The familons consist
of two adjoint Higgs fields ${\bf 45}_{(B-L)} (-1)$, 
${\bf 45}_X (-\frac{1}{2})$, and a singlet ${\bf 1}_H (-2)$.
The adjoint ${\bf 45}_{(B-L)}$ has VEV proportional to the generator $B-L$,
while ${\bf 45}_X$ has VEV proportional to the generator $X$, where
$SO(10) \supset SU(5) \times U(1)_X$. We will assume that all Yukawa
couplings are of order unity, that $\langle {\bf 45}_X \rangle \sim
M_{GUT}$, and that $\langle {\bf 45}_{(B-L)} \rangle \sim \epsilon M_{GUT}$,
and $\langle {\bf 1}_H \rangle \sim \delta M_{GUT}$, where $\delta \ll
\epsilon \ll 1$. 

With these fields, one has the couplings: (a) ${\bf 16}'_1 
\overline{{\bf 16}}'_1 \langle {\bf 45}_X \rangle$, ${\bf 16}'_2 
\overline{{\bf 16}}'_2 \langle {\bf 45}_X \rangle$, and 
${\bf 16}'_2 \overline{{\bf 16}}'_1 \langle {\bf 45}_{(B-L)} \rangle$, 
which give the matrices $M_f$; (b) ${\bf 16}_3 \overline{{\bf 16}}'_2
\langle {\bf 45}_{(B-L)} \rangle$ and 
${\bf 16}_2 \overline{{\bf 16}}'_2
\langle {\bf 1}_H \rangle$, which give the matrices $\hat{M}_f$;
(c) ${\bf 16}_3 {\bf 16}_3 \langle {\bf 10}_H \rangle$, which gives the
matrices $m_f$; and (d) ${\bf 16}_2 {\bf 16}'_1 \langle {\bf 10}_H
\rangle$, which gives the matrices $m'_f = m^{\prime \prime \prime T}_f$.
(Here the subscript $f$ stands for any of the types of fermion or
antifermion.)
We may then write the various mass matrices in the following form:
\begin{equation}
\begin{array}{cclccl}
m_f & = & \left( \begin{array}{ccc} 0 & 0 & 0 \\ 0 & 0 & 0 \\ 0 & 0 & 1
\end{array} \right) m_{U,D}, & & & \\ & & & & & \\
m'_f & = & m_f^{\prime \prime \prime T} = \left( \begin{array}{cc}
0 & 0 \\ x & 0 \\ 0 & 0 \end{array} \right) m_{U,D}, &
m^{\prime \prime}_f & = & \left( \begin{array}{cc} 0 & 0 \\ 0 & 0 \end{array}
\right), \\ & & & & & \\
M_f & = & \left( \begin{array}{cc} X_f  & 0 \\ (B-L)_f \epsilon y & 
X_f  \end{array}
\right) M, & 
\hat{M}_f & = & \left( \begin{array}{cc} 0 & 0 \\ 0 & \delta \\ 
0 & (B-L)_f \epsilon  
\end{array} \right) M. \end{array}
\end{equation}
\noindent
The weak-scale masses denoted $m_U$ and $m_D$ are proportional, respectively, 
to the VEVs $v_u$ and $v_d$. The mass $M$ is of order the unification scale.
$x,y \sim 1$. The foregoing allows us to write 
$Z_f \equiv \hat{M}_f M^{-1}_f$ as
\begin{equation}
Z_f = \left( \begin{array}{cc} 0 & 0 \\ -\frac{(B-L)_f}{X_f^2} 
\delta \epsilon y
& \frac{1}{X_f} \delta \\ - \frac{(B-L)_f^2}{X^2_f} \epsilon^2 y & 
\frac{(B-L)_f}{X_f} \epsilon \end{array} \right).
\end{equation}
\noindent
We shall neglect all terms that are subleading in $\delta$ and $\epsilon$.
That means that we may treat the matrices
$U_f$ and $V_f$ as being equal to the identity. The effective mass matrices 
$\overline{m}_f$ of the light states are then given by
$\overline{m}_f \cong m_f - m'_f Z_{f^c}^T - Z_f m_f^{\prime T}$. (See
Eqs. (6) and (25).) This gives
\begin{equation}
\overline{m}_f = \left( \begin{array}{ccc} 0 & 0 & 0 \\
0 & \left( \frac{(B-L)_f}{X_f^2} + \frac{(B-L)_{f^c}}{X_{f^c}^2}
\right) \delta \epsilon x y & \frac{(B-L)^2_{f^c}}{X^2_{f^c}} \epsilon^2 x y \\
0 & \frac{(B-L)^2_f}{X^2_f} \epsilon^2 x y & 1 \end{array} \right) m_{U,D}.
\end{equation}
\noindent
In particular, for the up quark, down quark, neutrino Dirac, and charged lepton
mass matrices, one has
\begin{equation}
\begin{array}{cclccl}
\overline{m}_u & = & \left( \begin{array}{ccc} 0 & 0 & 0 \\
0 & 0 & \frac{1}{9} \epsilon^2 x y \\
0 & \frac{1}{9} \epsilon^2 x y & 1 \end{array} \right) m_U, 
& \overline{m}_d & = &  \left( \begin{array}{ccc} 0 & 0 & 0 \\
0 & \frac{8}{27} \delta \epsilon x y & \frac{1}{81} \epsilon^2 x y \\
0 & \frac{1}{9} \epsilon^2 x y & 1 \end{array} \right) m_D \\ & & & & & \\
\overline{m}_{\nu} & = & \left( \begin{array}{ccc} 0 & 0 & 0 \\
0 & -\frac{16}{225} \delta \epsilon x y & \frac{1}{25} \epsilon^2 x y \\
0 & \frac{1}{9} \epsilon^2 x y & 1 \end{array} \right) m_U, 
& \overline{m}_{\ell} & = &  \left( \begin{array}{ccc} 0 & 0 & 0 \\
0 & \frac{8}{9} \delta \epsilon x y & \epsilon^2 x y \\
0 & \frac{1}{9} \epsilon^2 x y & 1 \end{array} \right) m_D \end{array}
\end{equation}
\noindent
The diagrams giving the 33, 23, 32, and 22 elements of $\overline{m}_{\nu}$
are shown in Figure 3. We see illustrated in Eq. (28) the general facts, 
noted earlier, that in unified
models the effective Dirac mass matrix of the neutrinos 
$\overline{m}_{\nu}$, given by Eq. (6), is related by the grand unified 
symmetries to the effective mass matrices of the quarks and charged leptons
and is hierarchical in form, like them. 

We are now in a position to determine the form of the mass matrix 
$M_{\nu \nu}$ of the light neutrinos. The model under consideration
falls in the class considered in section 3.1. Thus $M_{\nu \nu}$ is
given by Eq. (14). Since we are not assuming that $SU(2)$-triplet
Higgs fields acquire VEVs, we may set $\overline{\mu} =0$. If 
the matrix $\tilde{{\cal M}}$ vanishes, then as shown in section 3.1 the
term $m_* {\cal M}_*^{-1} m_*^T$ vanishes, and the expression for 
$M_{\nu \nu}$ reduces to the usual see-saw form. It should be emphasized
that this is not obvious by inspecting diagrams. For instance, if the
matrix ${\cal M}^{\prime \prime} \neq 0$, there are diagrams (shown in 
Fig. 4(a)) that do not involve the elements of $\tilde{{\cal M}}$ but that
nevertheless appear naively to give extra contributions to $M_{\nu \nu}$. 
However, our analysis shows (surprisingly) that such diagrams produce 
nothing beyond the usual see-saw terms.

Let us now look at some concrete cases. Suppose for simplicity that
${\cal M}'$ and ${\cal M}^{\prime \prime}$ vanish. Then, from the general
expressions for the barred matrices (see the equations in text after 
Eq. (11)), and neglecting terms higher order in the small quantities 
$\epsilon$ and $\delta$, one has that $\overline{{\cal M}}
= {\cal M}$, $\overline{{\cal M}'} = {\cal M} Z_{N^c}^*$, and
$\overline{{\cal M}^{\prime \prime}} = 0$. Now consider some simple
forms for ${\cal M}$. The easiest form 
to achieve in a model is skew-diagonal, as that can arise from the VEV of
a single $\overline{{\bf 126}}_H$ with family charge $-2$. (The
relevant terms are $({\bf 16}_1 {\bf 16}_3 + {\bf 16}_2 {\bf 16}_2)
\overline{{\bf 126}}_H$.) If we assume that ${\cal M} =
{\rm skewdiag} (A, 1, A) M_R$, with $A \sim 1$, Eq. (13) would give
\begin{equation}
M_{\nu \nu} = \frac{m^2_U}{M_R} \left( \frac{\epsilon x y}{9} \right)^2
\left( \begin{array}{ccc} 0 & 0 & 0 \\ 
0 & (\frac{16}{25})^2 \delta^2 & - \frac{16}{25} \delta \epsilon \\
0 & - \frac{16}{25} \delta \epsilon & \epsilon^2 \end{array}
\right).
\end{equation}
\noindent
Another simple possibility, in the sense of having
few parameters, is a diagonal form. If we assume that 
${\cal M} = {\rm diag} (B, A, 1) M_R$, with $A,B \sim 1$, Eq. (13) would give
\begin{equation}
M_{\nu \nu} = \frac{m^2_U}{M_R}
\left( \begin{array}{ccc} 0 & 0 & 0 \\ 
0 & O(\delta^2 \epsilon^2, \epsilon^4) & \frac{1}{25} \epsilon^2 x y \\
0 & \frac{1}{25} \epsilon^2 x y & 1 \end{array}
\right).
\end{equation}
\noindent
Note, in both cases, that the hierarchical structure of $\overline{m}_{\nu}$
leads to a hierarchical structure for $M_{\nu \nu}$ This can be avoided
if there is a strong hierarchy in the parameters of ${\cal M}$ that cancels
out the hierarchy in the Dirac mass matrix. (In that case some terms we dropped
because they were higher order in small parameters might have to be
retained as they could be multiplied by large quantities coming from
${\cal M}^{-1}$.) However, this requires some conspiracy between the
Dirac and Majorana mass matrices in the see-saw formula.

More interesting is the possibility that the extra term in Eq. (14)
could lead to a non-hierarchical pattern for $M_{\nu \nu}$.
Let us then compute $m_*$ and ${\cal M}_*$. We have already 
found that $\overline{m}_{\nu} = m_{\nu} + O(\delta, \epsilon)$. 
From Eq. (11), $\overline{m'}_{\nu} = m'_{\nu} + m_{\nu} Z^*_{N^c}$, 
neglecting terms higher order in small quantities. Therefore, from Eq. (15), 
and using the fact that $\overline{{\cal M}}^{-1} \overline{{\cal M}'} 
= Z^*_{N^c}$, one has $m_* = (m'_{\nu} + m_{\nu} Z^*_{N^c}) -
(m_{\nu} + O(\epsilon, \delta)) Z^*_{N^c}$. So $m_* = m'_{\nu}$ if we
neglect terms higher order in small quantities. 

From Eq. (16), ${\cal M}_* = - M_{N^c} \tilde{{\cal M}}^{-1} M_{N^c}^T
+ O(\epsilon^2, \delta^2)$. For $\tilde{{\cal M}}$
we choose the simple form $\tilde{{\cal M}} = 
{\rm diag}(\tilde{{\cal M}}_{11},\tilde{{\cal M}}_{22})$. 
Then, simply multiplying gives
\begin{equation}
\delta M_{\nu \nu} = m_* {\cal M}_*^{-1} m_*^T = \frac{m_U^2}{25 M^2}
(\tilde{{\cal M}}_{11} + \frac{y^2}{25} \tilde{{\cal M}}_{22}) 
\left( \begin{array}{ccc} 0 & 0 & 0 \\ 0 & 1 & 0 \\
0 & 0 & 0 \end{array} \right).
\end{equation}
\noindent
Note that this contribution to $M_{\nu \nu}$ is in no way hierarchically
suppressed. The small familon VEVs ($\langle {\bf 45}_{(B-L)} \rangle \sim
\epsilon M_{GUT}$, $\langle {\bf 1}_H \rangle \sim \delta M_{GUT}$)
that suppress some terms in the neutrino Dirac mass matrix (cf. Eq. (28)
and Fig. 3) do not enter this expression, as can be understood from its 
diagrammatic form, shown in Fig. 4(b).

We have been ignoring the masses of the first family. These arise
from an effective operator that is sixth order in a familon field
that is an adjoint of $SO(10)$ and has family charge $-\frac{1}{2}$. 
In Appendix A of Ref. \cite{adhrs}, this operator is shown to arise 
from integrating
out additional spinor-antispinor pairs (in fact, five of them). These
additional multiplets can can lead, in the
same way that we have been discussing, to large entries for the first family 
in $m_* {\cal M}_*^{-1} m_*^T$. 

\subsection{A model where ${\bf R}$ contains spinor-antispinor pairs, vectors, 
and singlets}

In the papers of Ref. \cite{abb} a very predictive $SO(10)$ 
models of quark and 
lepton masses is described. The fields that are integrated out to
produce the higher-dimension effective Yukawa operators that give mass
to the light families include spinors, vectors, and singlets. This model
therefore falls into the category studied in section 3.3. As in the previous
example studied, we shall for the sake of simplicity ignore the masses
of the first family and the fields that are integrated out to generate
them.

In the model of Ref. \cite{abb}, the quark and lepton content is the 
following, where
we indicate in parentheses the names of the neutrinos and antineutrinos
contained in each $SO(10)$ multiplet: ${\bf 16}_i (\nu_i, N^c_i)$, $i =
1,2,3$; ${\bf 16} (\nu'_1, N^{c \prime}_1)$; $\overline{{\bf 16}} 
(\overline{\nu}'_1, \overline{N^c}'_1)$; ${\bf 10} 
(\nu'_2, \overline{\nu}'_2)$; ${\bf 1}_i (S_i)$, $i=1,2,3$. (We have slightly
simplified the model: in the original there were two vectors of quarks and
leptons. We have chosen to identify them. This does not change the
model significantly.) The 
$SU(2)_L \times U(1)_Y$ breaking is done by a vector ${\bf 10}_H$ and a
spinor ${\bf 16}'_H$. The GUT-scale breaking is done by an adjoint
${\bf 45}_H$ whose VEV points in the $B-L$ direction, and by a spinor
${\bf 16}_H$ whose VEV is in the standard-model-singlet direction.

There are the following terms: (a) $M_{16} {\bf 16} \; \overline{{\bf 16}}
+ M_{10} {\bf 10} \; {\bf 10}$, which give the matrices $M_{\nu}$ and
$M_{N^c}$; (b) ${\bf 16}_3 \overline{{\bf 16}} \langle {\bf 45}_H \rangle +
{\bf 16}_2 {\bf 10} \langle {\bf 16}_H \rangle$, which give the matrices
$\hat{M}_{\nu}$ and $\hat{M}_{N^c}$; (c) ${\bf 16}_3 {\bf 16}_3 \langle
{\bf 10}_H \rangle$, which gives the matrix $m_{\nu}$; (d) ${\bf 16}_2
{\bf 16} \langle {\bf 10}_H \rangle$, which gives the matrices $m'_{\nu}$
and $m^{\prime \prime \prime}_{\nu}$; (e) $(M_1)_{ij} {\bf 1}_i \; {\bf 1}_j$, 
which gives the matrix $M_S$; (f) ${\bf 16}_i {\bf 1}_j \langle 
\overline{{\bf 16}}_H \rangle$, which gives the matrix $G$; and (optionally)
$\overline{{\bf 16}} \; {\bf 1}_i \langle {\bf 16}_H \rangle$, which
gives the matrix (in this case a $3 \times 1$ matrix) $\tilde{G}$. (See
Eq. (20) for the definitions.) 

It is simple to see that these matrices have the following forms:
\begin{equation}
\begin{array}{cclccl}
M_{\nu} & = & \left( \begin{array}{cc} M_{16} & 0 \\ 0 & M_{10} 
\end{array} \right), & M_{N^c} & = & \left( M_{16} \right), \\ & & & & & \\
\hat{M}_{\nu} & = & \left( \begin{array}{cc} 0 & 0 \\ 0 & \gamma \\
\beta & 0 \end{array} \right) M_{16}, & \hat{M}_{N^c} & = & \left(
\begin{array}{c} 0 \\ 0 \\ - \beta \end{array} \right) M_{16}, \\ & & & & & \\
m_{\nu} & = & \left( \begin{array}{ccc} 0 & 0 & 0 \\ 0 & 0 & 0 \\ 
0 & 0 & 1 \end{array} \right) m_U, & m^{\prime \prime}_{\nu} & = & 
\left( \begin{array}{c} 0 \\ 0 \end{array} \right), \\ & & & & & \\
m'_{\nu} & = & \left( \begin{array}{c} 0 \\ \alpha \\ 0 \end{array}
\right) m_U, & m^{\prime \prime \prime}_{\nu} & = & \left( \begin{array}{ccc}
0 & \alpha & 0 \\ 0 & 0 & 0 \end{array} \right) m_U. \end{array}
\end{equation}
\noindent
Then the $Z_f \equiv \hat{M}_f M^{-1}_f$ are given by
\begin{equation}
Z_{\nu} = \left( \begin{array}{cc} 0 & 0 \\ 0 & \gamma \frac{M_{16}}{M_{10}} \\
\beta & 0 \end{array} \right), \;\; Z_{N^c} = \left( \begin{array}{c}
0 \\ 0 \\ - \beta \end{array} \right).
\end{equation}
\noindent
We take the parameters $\beta$ and $\gamma$ to be small, and neglect terms
quadratic in them, so that the matrices $U_f$ and $V_f$ may be taken to 
be unity. From Eq. (6) it is simple to compute that $\overline{m}_{\nu}
\cong m_{\nu} - m'_{\nu} Z^T_{N^c} - Z_{\nu} m^{\prime \prime \prime}_{\nu}$,
or (defining $\epsilon \equiv \alpha \beta$):
\begin{equation}
\overline{m}_{\nu} = \left( \begin{array}{ccc} 0 & 0 & 0 \\
0 & 0 & \epsilon \\ 0 & - \epsilon & 1 \end{array} \right) m_U.
\end{equation}
\noindent
There are similar matrices for the up quarks, down quarks, and charged
leptons, but we will not bother to present them. 

The assumptions that led to Eq. (23) hold. Thus $M_{\nu \nu}$ has the
usual see-saw term, which will obviously be hierarchical, and possibly a
second term of the ``type III" form. Applying Eq. (24) to the present
model, we have $A = - \overline{m'}_{\nu} M_{16}^{-1} \tilde{G} G^{-1}$.
From Eq. (11), one has $\overline{m'}_{\nu} \cong m'_{\nu} + m_{\nu} Z_{N^c}$.
Defining $\tilde{G} G^{-1} \equiv (x_1, x_2, x_3)$, we then have altogether
\begin{equation}
A \cong \left( \begin{array}{ccc}
0 & 0 & 0 \\ -\alpha x_1 & - \alpha x_2 & - \alpha x_3 \\
\beta x_1 & \beta x_2 & \beta x_3 \end{array} \right) \frac{m_U}{M_{16}},
\end{equation}
\noindent
and thus, by Eq. (23),
\begin{equation}
\delta M_{\nu \nu} \cong \left( \begin{array}{ccc}
0 & 0 & 0 \\ 0 & - 2 \epsilon \alpha x_3 & 
- \alpha x_3 \\ 0 & - \alpha x_3 & 2 \epsilon \alpha^{-1} x_3 \end{array} 
\right) \frac{m_U^2}{M_{16}}.
\end{equation}

We see, again, that non-hierarchical contributions result. In particular,
in the usual see-saw term, the 23 and 32 elements are of order $\epsilon$,
whereas in this extra contribution these elements are not suppressed
($\alpha$ has not been assumed to be small).

\section{Conclusions}

We have given a fairly general analysis of neutrino mass in the context 
of $SO(10)$. Larger groups can be analyzed by decomposing to $SO(10)$.
We have seen that in grand unified models there can be additional contributions
to the mass matrix of the light neutrinos besides the usual ``type I"
see-saw term (or the ``type II" see-saw term arising from the
VEVs of triplet Higgs fields \cite{type2} and ``type III" see-saw term pointed 
out recently \cite{type3}). 

The additional contributions that we have discussed arise from 
integrating out the extra multiplets of fermions which typically exist 
at the unification scale in realistic models.  We have found that in certain 
special cases the light neutrino mass matrix does reduce to just the type I 
(and possibly also type II) term. These special cases include the 
following. (a) The extra quark and lepton 
multiplets at the GUT scale consist only of spinors and antispinors of
$SO(10)$ and there exist no superheavy mass terms of the form
$\overline{{\bf 16}} \; \overline{{\bf 16}} \langle {\bf 126}_H \rangle$. 
(b) The
extra quark and lepton multiplets consist only of spinors, antispinors,
and vectors of $SO(10)$, and there are neither superheavy mass terms of the
form $\overline{{\bf 16}} \; 
\overline{{\bf 16}} \langle {\bf 126}_H \rangle$, nor 
weak-scale masses of the form ${\bf 10} \; \overline{{\bf 16}} \langle 
\overline{{\bf 16}}_H \rangle$.
Except in special cases, however, there are non-type I extra terms. The
forms of these extra terms are given for various cases in Eqs. (14)-(16) and
Eq. (19). 

Typically, the Dirac mass matrix of the neutrinos that appears in the
type I see-saw term is just the effective Dirac mass neutrino matrix
$\overline{m}_{\nu}$ that is closely related by unified symmetries
to the effective mass matrices of the three observed families of up quarks,
down quarks, and charged leptons. Like those matrices, $\overline{m}_{\nu}$
is expected to be hierarchical in structure. Unless there is a ``conspiracy"
between the Dirac and Majorana mass matrices appearing in the type I see-saw
formula, one would therefore expect the light neutrino mass 
matrix $M_{\nu \nu}$ to be strongly hierarchical as well. However, the 
neutrino oscillation data implies that the neutrino masses are not 
strongly hierarchical, and
the solar and atmospheric angles are not small. This is somewhat of a
puzzle. 

The additional non-type I contributions that we have studied
in this paper might explain why $M_{\nu \nu}$ is not observed to
be strongly hierarchical. In fact, we have shown by examining two realistic
$SO(10)$ unified schemes that exist in the literature that the non-type I
contributions can indeed be large and lead to non-hierarchical patterns
for $M_{\nu \nu}$. As these examples show, especially the example discussed
in section 4.1, the family symmetries that are responsible in some models for 
the hierarchical ``textures" of the quark and charged lepton mass matrices,
and indeed of the neutrino Dirac mass matrix, do not in general lead to a
hierarchical form for the non-type I contributions to the neutrino mass matrix.

A significant point of our analysis is that the neutrino masses ``see" deeper
into the structure of the theory at the unification scale than do the
masses of the quarks and charged leptons. In fact, the neutrino masses
can see beyond the structure that leads the other masses to have a family 
hierarchy and thus itself be non-hierarchical. This last point has also
been made in the recent paper of Nir and Shadmi \cite{nirshadmi}, with which
we became familiar after completing the analysis in this paper.

The paper of Nir and Shadmi has certain obvious points of contact with the
present work, although the starting points are very different. The starting
point of Nir and Shadmi is family symmetry. They consider models of the 
Froggatt-Nielson type, in which the family hierarchies of the quarks
and charged leptons are controlled by a small parameter $\lambda_H$ that
is the ratio of the family-symmetry-breaking scale to the masses of the
vectorlike fermions that appear in Froggatt-Nielson models. They note that 
neutrino masses, on the other hand, being
Majorana rather than Dirac, involve an additional dimensionless parameter,
namely the ratio of the $(B-L)$-breaking scale to the masses of the
vectorlike fermions. Consequently, the neutrino mass matrix can have a
very different kind of hierarchy than the other types of fermion have.

Our starting point has been grand unified symmetry. The question we have
tried to answer is under what general conditions the usual (type I) 
see-saw formula is justified in grand unified theories, and when on the 
contrary this formula receives corrections. We have not paid attention 
in the general analysis of section 3 to the issue of family symmetry or 
the scale at which it is broken if it exists. 

The approach of the present paper and of Nir and Shadmi are thus, in a sense,
orthogonal. Froggatt-Nielson models need not be (and usually are not)
grand unified, and grand unified models need not be (and usually are not)
of the Froggatt-Nielson type. Nevertheless, a model can be both, as is the
example we have analyzed in section 4.1.

The key points that are common to our paper and that of Nir and Shadmi
is that vectorlike fermions typically exist in models that seek to explain
family hierarchies, and that the existence of these fermions can lead
to a non-hierarchical structure for the effective mass matrix of the three
light neutrinos.

\newpage

\noindent
{\bf\large Figure Captions}

\vspace{1cm}

\noindent
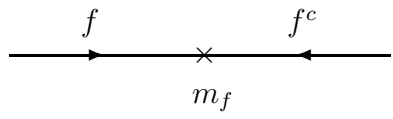
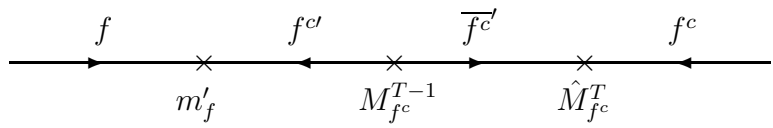
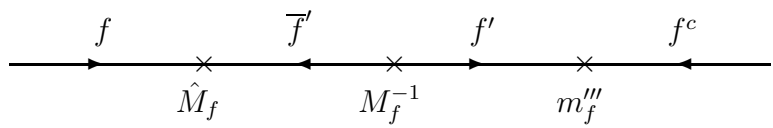
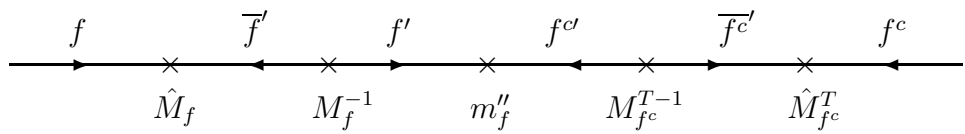
{\bf Fig 1.} Diagrams corresponding to the four terms of Eq. (6). 
$\overline{m}_f$ is the effective mass matrix of the light $f$, $f^c$
obtained by integrating out the superheavy multiplets $f + \overline{f}$,
and $f^c + \overline{f^c}$.

\vspace{0.2cm}

\noindent
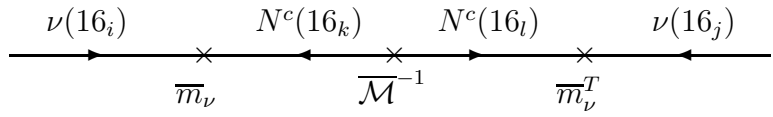
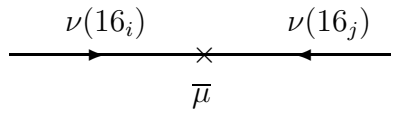
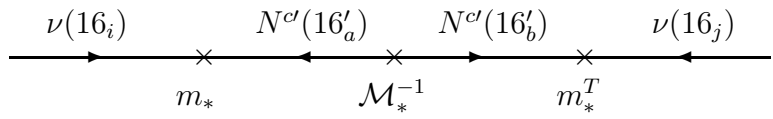
{\bf Fig 2.} Contributions to $M_{\nu \nu}$. (a) Usual ``type I" see-saw term.
(b) The ``type II" see-saw term, which comes from triplets Higgs fields. 
(c) The extra term that can exist if there are $\overline{{\bf 16}} \;
\overline{{\bf 16}} \; {\bf 126}_H$ terms. See Eqs. (14)-(16).

\vspace{0.2cm}

\noindent
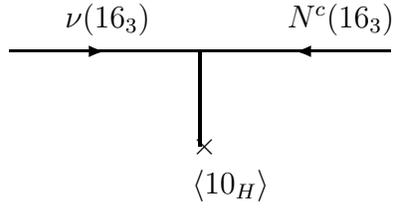
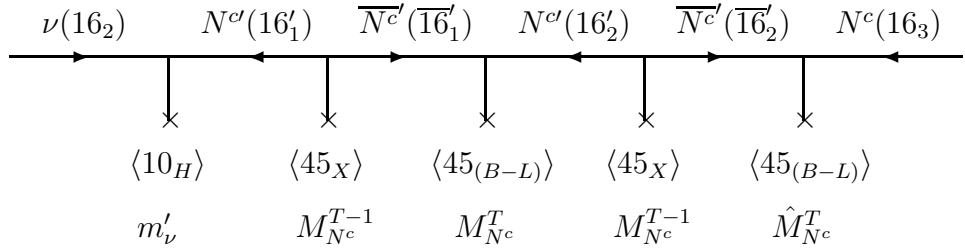
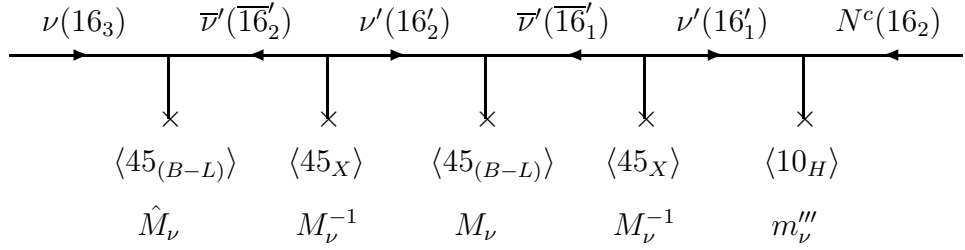
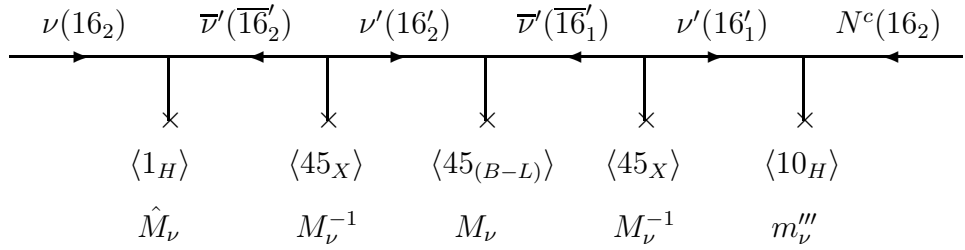
{\bf Fig. 3.} Diagrams that give the 33, 23, 32, and 22 elements of
$\overline{m}_{\nu}$ in the model discussed in section 4.1.

\vspace{0.2cm}

\noindent
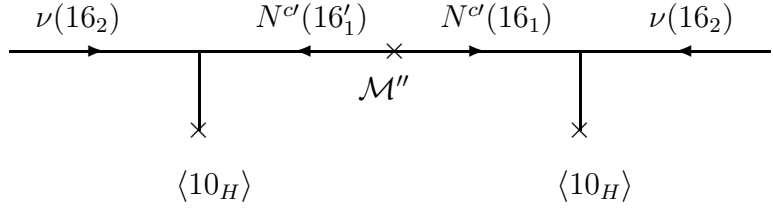
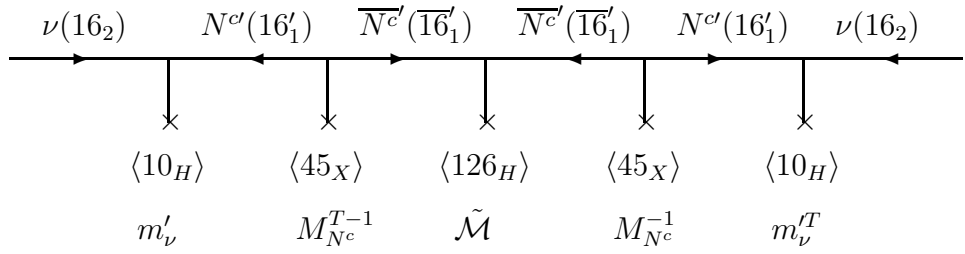
{\bf Fig 4.} (a) An apparent non-type-I-see-saw contribution to 
$M_{\nu \nu}$ in model of section 4.1, which actually vanishes 
if $\tilde{{\cal M}} = 0$.(b) A genuine non-type-I-see-saw 
contribution in that model.

\newpage

\begin{picture}(360,216)
\thicklines
\put(108,108){\vector(1,0){36}}
\put(144,108){\line(1,0){36}}
\put(180,108){\line(1,0){36}}
\put(252,108){\vector(-1,0){36}}
\put(177,105){$\times$}
\put(135,117){$f$}
\put(177,90){$m_f$}
\put(213,117){$f^c$}
\put(168,0){{\bf Fig. 1(a)}}
\end{picture}

\vspace{2cm}

\begin{picture}(360,216)
\thicklines
\put(36,108){\vector(1,0){36}}
\put(72,108){\line(1,0){36}}
\put(105,105){$\times$}
\put(108,108){\line(1,0){36}}
\put(180,108){\vector(-1,0){36}}
\put(177,105){$\times$}
\put(180,108){\vector(1,0){36}}
\put(216,108){\line(1,0){36}}
\put(249,105){$\times$}
\put(252,108){\line(1,0){36}}
\put(324,108){\vector(-1,0){36}}
\put(68,117){$f$}
\put(99,90){$m'_f$}
\put(141,117){$f^{c \prime}$}
\put(168,90){$M_{f^c}^{T-1}$}
\put(207,117){$\overline{f^c}'$}
\put(243,90){$\hat{M}_{f^c}^T$}
\put(285,117){$f^c$}
\put(168,0){{\bf Fig. 1(b)}}
\end{picture}

\newpage

\begin{picture}(360,216)
\thicklines
\put(36,108){\vector(1,0){36}}
\put(72,108){\line(1,0){36}}
\put(105,105){$\times$}
\put(108,108){\line(1,0){36}}
\put(180,108){\vector(-1,0){36}}
\put(177,105){$\times$}
\put(180,108){\vector(1,0){36}}
\put(216,108){\line(1,0){36}}
\put(249,105){$\times$}
\put(252,108){\line(1,0){36}}
\put(324,108){\vector(-1,0){36}}
\put(68,117){$f$}
\put(99,90){$\hat{M}_f$}
\put(141,117){$\overline{f}'$}
\put(168,90){$M_f^{-1}$}
\put(210,117){$f'$}
\put(243,90){$m^{\prime \prime \prime}_f$}
\put(285,117){$f^c$}
\put(168,0){{\bf Fig. 1(c)}}
\end{picture}

\vspace{2cm}

\begin{picture}(360,216)
\thicklines
\put(0,108){\vector(1,0){30}}
\put(30,108){\line(1,0){30}}
\put(60,108){\line(1,0){30}}
\put(120,108){\vector(-1,0){30}}
\put(120,108){\vector(1,0){30}}
\put(150,108){\line(1,0){30}}
\put(180,108){\line(1,0){30}}
\put(240,108){\vector(1,0){30}}
\put(240,108){\vector(-1,0){30}}
\put(270,108){\line(1,0){30}}
\put(300,108){\line(1,0){30}}
\put(360,108){\vector(-1,0){30}}
\put(56,105){$\times$}
\put(116,105){$\times$}
\put(176,105){$\times$}
\put(236,105){$\times$}
\put(296,105){$\times$}
\put(22,117){$f$}
\put(88,117){$\overline{f}'$}
\put(142,117){$f'$}
\put(202,117){$f^{c \prime}$}
\put(268,117){$\overline{f^c}'$}
\put(328,117){$f^c$}
\put(54,88){$\hat{M}_f$}
\put(114,88){$M_f^{-1}$}
\put(174,88){$m^{\prime \prime}_f$}
\put(225,88){$M_{f^c}^{T -1}$}
\put(294,88){$\hat{M}_{f^c}^T$}
\put(168,0){{\bf Fig. 1(d)}}
\end{picture}

\newpage

\begin{picture}(360,216)
\thicklines
\put(36,108){\vector(1,0){36}}
\put(72,108){\line(1,0){36}}
\put(105,105){$\times$}
\put(108,108){\line(1,0){36}}
\put(180,108){\vector(-1,0){36}}
\put(177,105){$\times$}
\put(180,108){\vector(1,0){36}}
\put(216,108){\line(1,0){36}}
\put(249,105){$\times$}
\put(252,108){\line(1,0){36}}
\put(324,108){\vector(-1,0){36}}
\put(50,117){$\nu(16_i)$}
\put(99,90){$\overline{m}_{\nu}$}
\put(129,117){$N^c(16_k)$}
\put(168,90){$\overline{{\cal M}}^{-1}$}
\put(198,117){$N^c(16_l)$}
\put(243,90){$\overline{m}_{\nu}^T$}
\put(279,117){$\nu(16_j)$}
\put(168,0){{\bf Fig. 2(a)}}
\end{picture}

\vspace{2cm}

\begin{picture}(360,216)
\thicklines
\put(108,108){\vector(1,0){36}}
\put(144,108){\line(1,0){36}}
\put(180,108){\line(1,0){36}}
\put(252,108){\vector(-1,0){36}}
\put(177,105){$\times$}
\put(129,117){$\nu(16_i)$}
\put(177,90){$\overline{\mu}$}
\put(213,117){$\nu(16_j)$}
\put(168,0){{\bf Fig. 2(b)}}
\end{picture}

\newpage

\begin{picture}(360,216)
\thicklines
\put(36,108){\vector(1,0){36}}
\put(72,108){\line(1,0){36}}
\put(105,105){$\times$}
\put(108,108){\line(1,0){36}}
\put(180,108){\vector(-1,0){36}}
\put(177,105){$\times$}
\put(180,108){\vector(1,0){36}}
\put(216,108){\line(1,0){36}}
\put(249,105){$\times$}
\put(252,108){\line(1,0){36}}
\put(324,108){\vector(-1,0){36}}
\put(50,117){$\nu(16_i)$}
\put(99,90){$m_*$}
\put(129,117){$N^{c \prime}(16'_a)$}
\put(168,90){${\cal M}_*^{-1}$}
\put(198,117){$N^{c \prime}(16'_b)$}
\put(243,90){$m_*^T$}
\put(279,117){$\nu(16_j)$}
\put(168,0){{\bf Fig. 2(c)}}
\end{picture}

\newpage

\begin{picture}(360,216)
\thicklines
\put(108,108){\vector(1,0){36}}
\put(144,108){\line(1,0){36}}
\put(180,108){\line(1,0){36}}
\put(252,108){\vector(-1,0){36}}
\put(180,72){\line(0,1){36}}
\put(177,69){$\times$}
\put(129,117){$\nu(16_3)$}
\put(177,54){$\langle 10_H \rangle$}
\put(213,117){$N^c(16_3)$}
\put(168,0){{\bf Fig. 3(a)}}
\end{picture}

\vspace{2cm}

\begin{picture}(360,216)
\thicklines
\put(0,108){\vector(1,0){30}}
\put(30,108){\line(1,0){30}}
\put(60,108){\line(1,0){30}}
\put(120,108){\vector(-1,0){30}}
\put(120,108){\vector(1,0){30}}
\put(150,108){\line(1,0){30}}
\put(180,108){\line(1,0){30}}
\put(240,108){\vector(1,0){30}}
\put(240,108){\vector(-1,0){30}}
\put(270,108){\line(1,0){30}}
\put(300,108){\line(1,0){30}}
\put(360,108){\vector(-1,0){30}}
\put(60,84){\line(0,1){24}}
\put(120,84){\line(0,1){24}}
\put(180,84){\line(0,1){24}}
\put(240,84){\line(0,1){24}}
\put(300,84){\line(0,1){24}}
\put(56,81){$\times$}
\put(116,81){$\times$}
\put(176,81){$\times$}
\put(236,81){$\times$}
\put(296,81){$\times$}
\put(12,117){$\nu(16_2)$}
\put(72,117){$N^{c \prime} (16'_1)$}
\put(132,117){$\overline{N^c}'(\overline{16}'_1)$}
\put(192,117){$N^{c \prime} (16'_2)$}
\put(252,117){$\overline{N^c}' (\overline{16}'_2)$}
\put(312,117){$N^c (16_3)$}
\put(45,64){$\langle 10_H \rangle$}
\put(105,64){$\langle 45_X \rangle$}
\put(159,64){$\langle 45_{(B-L)} \rangle$}
\put(225,64){$\langle 45_X \rangle$}
\put(279,64){$\langle 45_{(B-L)} \rangle$}
\put(48,40){$m'_{\nu}$}
\put(108,40){$M_{N^c}^{T -1}$}
\put(168,40){$M_{N^c}^T$}
\put(228,40){$M_{N^c}^{T -1}$}
\put(288,40){$\hat{M}_{N^c}^T$}
\put(168,0){{\bf Fig. 3(b)}}
\end{picture}

\newpage 

\begin{picture}(360,216)
\thicklines
\put(0,108){\vector(1,0){30}}
\put(30,108){\line(1,0){30}}
\put(60,108){\line(1,0){30}}
\put(120,108){\vector(-1,0){30}}
\put(120,108){\vector(1,0){30}}
\put(150,108){\line(1,0){30}}
\put(180,108){\line(1,0){30}}
\put(240,108){\vector(1,0){30}}
\put(240,108){\vector(-1,0){30}}
\put(270,108){\line(1,0){30}}
\put(300,108){\line(1,0){30}}
\put(360,108){\vector(-1,0){30}}
\put(60,84){\line(0,1){24}}
\put(120,84){\line(0,1){24}}
\put(180,84){\line(0,1){24}}
\put(240,84){\line(0,1){24}}
\put(300,84){\line(0,1){24}}
\put(56,81){$\times$}
\put(116,81){$\times$}
\put(176,81){$\times$}
\put(236,81){$\times$}
\put(296,81){$\times$}
\put(12,117){$\nu(16_3)$}
\put(72,117){$\overline{\nu}' (\overline{16}'_2)$}
\put(132,117){$\nu'(16'_2)$}
\put(192,117){$\overline{\nu}' (\overline{16}'_1)$}
\put(252,117){$\nu' (16'_1)$}
\put(312,117){$N^c (16_2)$}
\put(39,64){$\langle 45_{(B-L)} \rangle$}
\put(105,64){$\langle 45_X \rangle$}
\put(159,64){$\langle 45_{(B-L)} \rangle$}
\put(225,64){$\langle 45_X \rangle$}
\put(285,64){$\langle 10_H \rangle$}
\put(48,40){$\hat{M}_{\nu}$}
\put(108,40){$M_{\nu}^{-1}$}
\put(168,40){$M_{\nu}$}
\put(228,40){$M_{\nu}^{-1}$}
\put(288,40){$m^{\prime \prime \prime}_{\nu}$}
\put(168,0){{\bf Fig. 3(c)}}
\end{picture}

\vspace{2cm}

\begin{picture}(360,216)
\thicklines
\put(0,108){\vector(1,0){30}}
\put(30,108){\line(1,0){30}}
\put(60,108){\line(1,0){30}}
\put(120,108){\vector(-1,0){30}}
\put(120,108){\vector(1,0){30}}
\put(150,108){\line(1,0){30}}
\put(180,108){\line(1,0){30}}
\put(240,108){\vector(1,0){30}}
\put(240,108){\vector(-1,0){30}}
\put(270,108){\line(1,0){30}}
\put(300,108){\line(1,0){30}}
\put(360,108){\vector(-1,0){30}}
\put(60,84){\line(0,1){24}}
\put(120,84){\line(0,1){24}}
\put(180,84){\line(0,1){24}}
\put(240,84){\line(0,1){24}}
\put(300,84){\line(0,1){24}}
\put(56,81){$\times$}
\put(116,81){$\times$}
\put(176,81){$\times$}
\put(236,81){$\times$}
\put(296,81){$\times$}
\put(12,117){$\nu(16_2)$}
\put(72,117){$\overline{\nu}' (\overline{16}'_2)$}
\put(132,117){$\nu'(16'_2)$}
\put(192,117){$\overline{\nu}' (\overline{16}'_1)$}
\put(252,117){$\nu' (16'_1)$}
\put(312,117){$N^c (16_2)$}
\put(45,64){$\langle 1_H \rangle$}
\put(105,64){$\langle 45_X \rangle$}
\put(159,64){$\langle 45_{(B-L)} \rangle$}
\put(225,64){$\langle 45_X \rangle$}
\put(285,64){$\langle 10_H \rangle$}
\put(48,40){$\hat{M}_{\nu}$}
\put(108,40){$M_{\nu}^{-1}$}
\put(168,40){$M_{\nu}$}
\put(228,40){$M_{\nu}^{-1}$}
\put(288,40){$m^{\prime \prime \prime}_{\nu}$}
\put(168,0){{\bf Fig. 3(d)}}
\end{picture}

\newpage

\begin{picture}(360,216)
\thicklines
\put(36,108){\vector(1,0){36}}
\put(72,108){\line(1,0){36}}
\put(108,78){\line(0,1){30}}
\put(103,75){$\times$}
\put(108,108){\line(1,0){36}}
\put(180,108){\vector(-1,0){36}}
\put(177,105){$\times$}
\put(180,108){\vector(1,0){36}}
\put(216,108){\line(1,0){36}}
\put(252,78){\line(0,1){30}}
\put(247,75){$\times$}
\put(252,108){\line(1,0){36}}
\put(324,108){\vector(-1,0){36}}
\put(46,117){$\nu (16_2)$}
\put(99,54){$\langle 10_H \rangle$}
\put(129,117){$N^{c \prime}(16'_1)$}
\put(168,90){${\cal M}^{\prime \prime}$}
\put(199,117){$N^{c \prime}(16_1)$}
\put(243,54){$\langle 10_H \rangle$}
\put(278,117){$\nu(16_2)$}
\put(168,0){{\bf Fig. 4(a)}}
\end{picture}

\vspace{2cm}

\begin{picture}(360,216)
\thicklines
\put(0,108){\vector(1,0){30}}
\put(30,108){\line(1,0){30}}
\put(60,108){\line(1,0){30}}
\put(120,108){\vector(-1,0){30}}
\put(120,108){\vector(1,0){30}}
\put(150,108){\line(1,0){30}}
\put(180,108){\line(1,0){30}}
\put(240,108){\vector(1,0){30}}
\put(240,108){\vector(-1,0){30}}
\put(270,108){\line(1,0){30}}
\put(300,108){\line(1,0){30}}
\put(360,108){\vector(-1,0){30}}
\put(60,84){\line(0,1){24}}
\put(120,84){\line(0,1){24}}
\put(180,84){\line(0,1){24}}
\put(240,84){\line(0,1){24}}
\put(300,84){\line(0,1){24}}
\put(56,81){$\times$}
\put(116,81){$\times$}
\put(176,81){$\times$}
\put(236,81){$\times$}
\put(296,81){$\times$}
\put(12,117){$\nu(16_2)$}
\put(72,117){$N^{c \prime} (16'_1)$}
\put(132,117){$\overline{N^c}'(\overline{16}'_1)$}
\put(192,117){$\overline{N^c}'(\overline{16}'_1)$}
\put(252,117){$N^{c \prime} (16'_1)$}
\put(312,117){$\nu(16_2)$}
\put(45,64){$\langle 10_H \rangle$}
\put(105,64){$\langle 45_X \rangle$}
\put(162,64){$\langle 126_H \rangle$}
\put(225,64){$\langle 45_X \rangle$}
\put(285,64){$\langle 10_H \rangle$}
\put(48,40){$m'_{\nu}$}
\put(108,40){$M_{N^c}^{T -1}$}
\put(168,40){$\tilde{{\cal M}}$}
\put(228,40){$M_{N^c}^{-1}$}
\put(288,40){$m_{\nu}^{\prime T}$}
\put(168,0){{\bf Fig. 4(b)}}
\end{picture}

\end{document}